\documentclass[11pt]{article}
\usepackage[centertags]{amsmath}
\usepackage{amsfonts}
\usepackage{amssymb}
\usepackage{amsthm}
\usepackage{newlfont}
\usepackage{graphicx}
\usepackage{multirow}
\usepackage{rotating}
\usepackage{epstopdf}
\usepackage{lscape}

\theoremstyle{definition}

\theoremstyle{remark}

\theoremstyle{theorem}

\newtheorem{theorem}{Theorem}

\begin{document}
\thispagestyle{empty}

%
%

\title{Solving a non-linear model of HIV infection for CD$4^+$T cells by combining Laplace transformation and Homotopy analysis
method}
\author{ Samad Noeiaghdam$^{1,}$
\footnote{Corresponding author, E-mail addresses:
s.noeiaghdam.sci@iauctb.ac.ir; samadnoeiaghdam@gmail.com}
~~~~~~~~~~Emran Khoshrouye Ghiasi$^{2,}$ \footnote{E-mail address:
khoshrou@yahoo.com }
}
  \date{}
 \maketitle

\begin{center}
\scriptsize{ $^1$
Department of Mathematics, Central Tehran Branch, Islamic Azad University, Tehran, Iran. 

$^2$Young Researchers and Elite Club, Mashhad Branch, Islamic Azad
University, Mashhad, Iran.\\


}
\end{center}
\begin{abstract}
The aim of this paper is to find the approximate solution of HIV
infection model of CD$4^+$T  cells. For this reason, the homotopy
analysis transform method (HATM) is applied. The presented method is
combination of traditional homotopy analysis method (HAM) and the
Laplace transformation. The convergence of presented method is
discussed by preparing a theorem which shows the capabilities of
method. The numerical results are shown for different values of
iterations. Also, the regions of convergence are demonstrated by
plotting several $\hbar$-curves. Furthermore in order to show the
efficiency and accuracy of method, the residual error for different
iterations are presented.

 \vspace{.5cm}{\it keywords:} Homotopy analysis method \and Laplace transformation \and
System of non-linear differential equations \and HIV infection model
of CD$4^+$T  cells.\\

{\it MSC codes (2000):} 92B05, 65L99, 44A10.
\end{abstract}
\section{Introduction}
In last decades, many natural phenomenons have been described by
various nonlinear mathematical models such as the human
immunodeficiency virus (HIV) \cite{1,12,15,16,17,18}, mathematical
model of smoking habit \cite{r5,r4,r6}, model of malaria
transmission \cite{r7}, model of computer viruses
\cite{r9,r10,r8,r11} and many mathematical models that have direct
role in our life. Perelson in 1989 presented a model for infection
of human immune system by HIV which was based on the three variables
\cite{r1}. Then Perelson et al. modified and generalized this model
to new model with four variables by a system of non-linear ordinary
differential equations \cite{r2}. In recent years, several
applicable models have been presented based on the Perelson's models
\cite{r3}. Also, many numerical and semi-analytical methods have
been applied to solve the vivo dynamics of $T$-cell and HIV
interaction. The aim of this paper is to solve the Culshaw and
Ruan's model \cite{r3} for HIV infection of CD4$^+$T-cells as
\begin{equation}\label{1}
\begin{array}{l}
\displaystyle  \frac{d\mathcal{T}(c)}{dc} = s - \omega_\mathcal{T} \mathcal{T}(c) + g \mathcal{T}(c) \left(1-\frac{\mathcal{I}(c)+\mathcal{T}(c)}{\mathcal{T}_{\max}}\right) - \alpha \mathcal{V} (c) \mathcal{T}(c), \\
   \\
\displaystyle \frac{d\mathcal{I}(c)}{dc} = \beta \mathcal{V}(c) \mathcal{T}(c) - \omega_I \mathcal{I}(c),\\
 \\
\displaystyle \frac{d\mathcal{V}(c)}{dc} = n \omega_b \mathcal{I}(c)
- \alpha \mathcal{V}(c) \mathcal{T}(c) - \omega_\mathcal{V}
\mathcal{V}(c),
\end{array}
\end{equation}
with the initial conditions
\begin{equation}\label{2}
\mathcal{T}(0) = \mathcal{T}_0(c) = \mathcal{T}_0, \mathcal{I}(0) =
\mathcal{I}_0(c) = \mathcal{I}_0, \mathcal{V}(0) = \mathcal{V}_0(c)
= \mathcal{V}_0.
\end{equation}

Functions $\mathcal{T} (c), \mathcal{I}(c),$ and $\mathcal{V}(c)$
show the concentration of susceptible CD$4^+$T cells, CD$4^+$T cells
infected by the HIV viruses and free HIV virus particles in the
blood, respectively. List of parameters are presented in Table
\ref{t0}.

\begin{table}[h]
\caption{ List of parameters and functions. }\label{t0}
 \centering
\scalebox{0.9}{
\begin{tabular}{|c|l|l|}
\hline
 Parameters &  ~~~~~~~~~~~~~~~~~~~~~~~~~Meaning   & Values      \\
 $\&$ Functions &&\\
    \hline
   $\mathcal{T}$ & Uninfected $CD4^+$ T-cell concentration &$\mathcal{T}_0 = 1000$\\
   $\mathcal{I}$ &Infected $CD4^+$ T-cell concentration&$\mathcal{I}_0 = 0$\\
   $\mathcal{V}$ & Concentration of HIV RNA &$\mathcal{V}_0 = 0.001$\\
   $\omega_\mathcal{T}$  &Natural death rate of $CD4^+$ T-cell concentration&0.02\\
   $\omega_I$  &Blanket death rate of infected $CD4^+$ T-cells&0.26\\
   $\omega_b $ &Lytic death rate for infected cells&0.24\\
   $\omega_\mathcal{V} $ &Death rate of free virus&2.4\\
   $\alpha $ &Rate $CD4^+$ T-cells become infected with virus&$2.4 \times 10^{-5}$\\
   $\beta$  &Rate infected cells become active&$2 \times 10^{-5}$\\
    $g$ &Growth rate of $CD4^+$ T-cell concentration&0.03\\
    $n$ &Number of virion produced by infected $CD4^+$ T-cells&500\\
    $\mathcal{T}_{\max}$ &Maximal concentration of $CD4^+$ T-cells&1500\\
    $s$ &Source term for uninfected $CD4^+$ T-cells&10\\
 \hline
 \end{tabular}
 }
\end{table}

Solving problem (\ref{1}) has been studied by many authors. Several
mathematical methods such as the Adomian decomposition method (ADM)
\cite{1,2,12}, HAM \cite{9}, variational iteration method (VIM)
\cite{10}, collocation method (CM) \cite{7,13,14,16,17},
differential transform method (DTM) \cite{8,15,19} and other
\cite{18} were applied to solve the system of Eqs. (\ref{1}).

The HAM \cite{n8,n9,n10,n11,n12,n13,4,6,r12} is an important and
useful method to solve many problems
\cite{em1,em2,em3,em4,em5,em6,em7}. The solution of this method
contains the optimal convergence control parameter to find the
region of convergence. Also this method depends on the auxiliary
function and linear operator to find the solution. Choosing the
proper function and operator can be leaded to the solution quickly.
These are the important capabilities of HAM that the other
traditional methods do not have these advantages.

In this paper, HAM is combined with the Laplace transformation
\cite{r14,r17,r15,r16,r13} to construct a new and robust method
which is called the homotopy analysis transform method (HATM)
\cite{5,n6,n7,11}. This method applied to solve many problems
arising in engineering and many sciences \cite{r20,r18,r19}. In this
study, by using the HATM we try to solve the non-linear system of
Eqs. (\ref{1}) and the numerical solutions for $N=5,10,15$ are
obtained. The convergence theorem warranties the homotopy analysis
transform method theoretically. Several $h$-curves are plotted to
show the regions of convergence. Also, in order to show the
efficiency and accuracy of HATM, the residual error functions are
applied for different values based on the regions of convergence.

\section{Homotopy analysis transform method}
In order to apply the HATM for solving system of Eqs. (\ref{1}), the
Laplace transformation $\mathcal{L}$ is used as follows
\begin{equation}\label{3}
\begin{array}{ll}
\displaystyle  \mathcal{L} \bigg[\mathcal{T}(c)\bigg] &\displaystyle
= \frac{\mathcal{T}(0)}{z} + \frac{\mathcal{L} \bigg[s\bigg]}{z} +
\frac{(g-\omega_\mathcal{T})}{z} \mathcal{L}
\bigg[\mathcal{T}(c)\bigg]
- \frac{g}{z \mathcal{T}_{\max}}  \mathcal{L} \bigg[\mathcal{T}^2(c)\bigg]  \\
\\& \displaystyle - \frac{g}{z \mathcal{T}_{\max}} \mathcal{L} \bigg[\mathcal{T}(c) \mathcal{I}(c)\bigg] - \frac{\alpha}{z} \mathcal{L} \bigg[\mathcal{V}(c) \mathcal{T}(c)\bigg], \\
   \\
\displaystyle \mathcal{L} \bigg[\mathcal{I}(c)\bigg] &\displaystyle = \frac{\mathcal{I}(0)}{z} +  \frac{\beta}{z} \mathcal{L} \bigg[\mathcal{V}(c) \mathcal{T}(c) \bigg]  - \frac{\omega_\mathcal{I}}{z} \mathcal{L} \bigg[\mathcal{I}(c)\bigg],\\
 \\
\displaystyle \mathcal{L} \bigg[\mathcal{V}(c)\bigg]& \displaystyle
= \frac{\mathcal{V}(0)}{z} + \frac{n \omega_b}{z} \mathcal{L}
\bigg[\mathcal{I}(c)\bigg] - \frac{\alpha}{z} \mathcal{L}\bigg[
\mathcal{V}(c) \mathcal{T}(c)\bigg] - \frac{\omega_\mathcal{V}}{z}
\mathcal{L}\bigg[\mathcal{V}(c)\bigg].
\end{array}
\end{equation}
According to \cite{n1,n2,n3,n4,n5} we define the following
non-linear operators $N_1, N_2$ and $N_3$ as
\begin{equation}\label{4}
\begin{array}{l}
\displaystyle  N_1 \bigg[\mathcal{T}(c;\eta), \mathcal{I}(c;\eta),
\mathcal{V}(c;\eta)\bigg] \\
\\
\displaystyle  = \mathcal{L} \bigg[\mathcal{T}(c;\eta)\bigg]
\displaystyle - \frac{\mathcal{T}(0)}{z} - \frac{\mathcal{L}
\bigg[s\bigg]}{z}
 -
\frac{(g-\omega_\mathcal{T})}{z} \mathcal{L}
\bigg[\mathcal{T}(c;\eta)\bigg]+ \frac{g}{z \mathcal{T}_{\max}}
\mathcal{L} \bigg[\mathcal{T}^2(c;\eta)\bigg]\\
\\
\displaystyle + \frac{g}{z \mathcal{T}_{\max}} \mathcal{L} \bigg[\mathcal{T}(c;\eta) \mathcal{I}(c;\eta)\bigg]  + \frac{\alpha}{z} \mathcal{L} \bigg[\mathcal{V}(c;\eta) \mathcal{T}(c;\eta)\bigg], \\
\\
\displaystyle N_2 \bigg[\mathcal{T}(c;\eta), \mathcal{I}(c;\eta),
\mathcal{V}(c;\eta)\bigg] \\
\\ \displaystyle   = \mathcal{L} \bigg[\mathcal{I}(c;\eta)\bigg]  \displaystyle -
\frac{\mathcal{I}(0)}{z} -  \frac{\beta}{z} \mathcal{L}
\bigg[\mathcal{V}(c;\eta) \mathcal{T}(c;\eta) \bigg]
 + \frac{\omega_\mathcal{I}}{z} \mathcal{L} \bigg[\mathcal{I}(c;\eta)\bigg],\\
 \\
\displaystyle N_3 \bigg[\mathcal{T}(c;\eta), \mathcal{I}(c;\eta),
\mathcal{V}(c;\eta)\bigg] \\
\\
\displaystyle  = \mathcal{L} \bigg[\mathcal{V}(c;\eta)\bigg]
\displaystyle  - \frac{\mathcal{V}(0)}{z} - \frac{n \omega_b}{z}
\mathcal{L} \bigg[\mathcal{I}(c;\eta)\bigg]   + \frac{\alpha}{z}
\mathcal{L}\bigg[ \mathcal{V}(c;\eta) \mathcal{T}(c;\eta)\bigg] +
\frac{\omega_\mathcal{V}}{z}
\mathcal{L}\bigg[\mathcal{V}(c;\eta)\bigg],
\end{array}
\end{equation}
where $\eta \in [0,1]$ is an imbedding parameter. Now the following
homotopy maps can be constructed as
\begin{equation}\label{5}
 \begin{array}{l}
 \mathcal{H}_1 \left(\tilde{\mathcal{T}}(c;\eta),\tilde{\mathcal{I}}(c;\eta),\tilde{\mathcal{V}}(c;\eta)\right)\\
 \\
  = (1-\eta) L_1\left[\tilde{\mathcal{T}}(c;\eta)-\mathcal{T}_0(c)\right] - \eta \hbar H_1(c) N_1 \left[\tilde{\mathcal{T}}(c;\eta), \tilde{\mathcal{I}}(c;\eta), \tilde{\mathcal{V}}(c;\eta)\right],\\
 \\
 \mathcal{H}_2 \left(\tilde{\mathcal{T}}(c;\eta),\tilde{\mathcal{I}}(c;\eta),\tilde{\mathcal{V}}(c;\eta)\right)\\
 \\
  = (1-\eta) L_2\left[\tilde{\mathcal{I}}(c;\eta)-\mathcal{I}_0(c)\right] - \eta \hbar H_2(c) N_2 \left[\tilde{\mathcal{T}}(c;\eta), \tilde{\mathcal{I}}(c;\eta), \tilde{\mathcal{V}}(c;\eta)\right],\\
\\
\mathcal{H}_3 \left(\tilde{\mathcal{T}}(c;\eta),\tilde{\mathcal{I}}(c;\eta),\tilde{\mathcal{V}}(c;\eta)\right)\\
\\
 = (1-\eta) L_3\left[\tilde{\mathcal{V}}(c;\eta)-\mathcal{V}_0(c)\right] - \eta \hbar H_3(c) N_3 \left[\tilde{\mathcal{T}}(c;\eta), \tilde{\mathcal{I}}(c;\eta), \tilde{\mathcal{V}}(c;\eta)\right],\\
 \end{array}
\end{equation}
where $\hbar$ is a nonzero auxiliary parameter, $H_1, H_2, H_3$ are
auxiliary functions and $N_1, N_2, N_3$ are non-linear operators.
When
$$
\begin{array}{l}
  \mathcal{H}_1
\left(\tilde{\mathcal{T}}(c;\eta),\tilde{\mathcal{I}}(c;\eta),\tilde{\mathcal{V}}(c;\eta)\right)
= \mathcal{H}_2
\left(\tilde{\mathcal{T}}(c;\eta),\tilde{\mathcal{I}}(c;\eta),\tilde{\mathcal{V}}(c;\eta)\right) \\
   \\
  = \mathcal{H}_3
\left(\tilde{\mathcal{T}}(c;\eta),\tilde{\mathcal{I}}(c;\eta),\tilde{\mathcal{V}}(c;\eta)\right)
= 0,
\end{array}
$$
the zero order deformation equations are obtained
 \begin{equation}\label{6}
 \begin{array}{l}
 (1-\eta) L_1\left[\tilde{\mathcal{T}}(c;\eta)-\mathcal{T}_0(c)\right] - \eta \hbar H_1(c) N_1 \left[\tilde{\mathcal{T}}(c;\eta), \tilde{\mathcal{I}}(c;\eta), \tilde{\mathcal{V}}(c;\eta)\right]=0,\\
 \\
 (1-\eta) L_2\left[\tilde{\mathcal{I}}(c;\eta)-\mathcal{I}_0(c)\right] - \eta \hbar H_2(c) N_2 \left[\tilde{\mathcal{T}}(c;\eta), \tilde{\mathcal{I}}(c;\eta), \tilde{\mathcal{V}}(c;\eta)\right]=0,\\
\\
 (1-\eta) L_3\left[\tilde{\mathcal{V}}(c;\eta)-\mathcal{V}_0(c)\right] - \eta \hbar H_3(c) N_3 \left[\tilde{\mathcal{T}}(c;\eta), \tilde{\mathcal{I}}(c;\eta), \tilde{\mathcal{V}}(c;\eta)\right]=0.\\
 \end{array}
 \end{equation}
Also, for $\eta=0$ and $\eta=1$ the homotopy equations lead to
$$
 \begin{array}{lll}
  \tilde{\mathcal{T}}(c;0) = \mathcal{T}_0(c), & ~~~~~ &  \tilde{\mathcal{T}}(c;1) = \mathcal{T}(c), \\
  \\
  \tilde{\mathcal{I}}(c;0) = \mathcal{I}_0(c), & ~~~~~ &  \tilde{\mathcal{I}}(c;1) = \mathcal{\mathcal{I}}(c), \\
  \\
  \tilde{\mathcal{V}}(c;0) = \mathcal{V}_0(c), & ~~~~~ &  \tilde{\mathcal{V}}(c;1) = \mathcal{V}(c). \\
    \end{array}
$$
Therefore, when $\eta$ changes from $0$ to $1$ the initial functions
$\mathcal{T}_0(c), \mathcal{I}_0(c), \mathcal{V}_0(c)$ lead to the
exact solutions $\mathcal{T}(c), \mathcal{I}(c), \mathcal{V}(c)$. By
applying the Taylor series for $\tilde{\mathcal{T}}(c;\eta),
\tilde{\mathcal{I}}(c;\eta), \tilde{\mathcal{V}}(c;\eta)$ with
respect to $\eta$ we get
\begin{equation}\label{7}
\begin{array}{l}
\displaystyle \tilde{\mathcal{T}}(c;\eta) = \mathcal{T}_0(c) + \sum_{d=1}^\infty \mathcal{T}_d(c) \eta^d,\\
\\
\displaystyle \tilde{\mathcal{I}}(c;\eta)= \mathcal{I}_0(c) + \sum_{d=1}^\infty \mathcal{I}_d(c) \eta^d,\\
 \\
\displaystyle  \tilde{\mathcal{V}}(c;\eta)= \mathcal{V}_0(c) +
\sum_{d=1}^\infty \mathcal{V}_d(c) \eta^d,
\end{array}
\end{equation}
where
$$
\begin{array}{lll}
\displaystyle \mathcal{T}_d = \frac{1}{d!} \frac{\partial^d
\tilde{\mathcal{T}}(c;\eta) }{\partial \eta^d}\bigg|_{\eta=0}, &
\displaystyle \mathcal{I}_d = \frac{1}{d!} \frac{\partial^d
\tilde{\mathcal{I}}(c;\eta) }{\partial \eta^d}\bigg|_{\eta=0}, &
\displaystyle \mathcal{V}_d = \frac{1}{d!} \frac{\partial^d  \tilde{\mathcal{V}}(c;\eta) }{\partial \eta^d}\bigg|_{\eta=0}.  \\
\end{array}
$$
For proper value, operator and functions of $\hbar, L, H_1(c),
H_2(c), H_3(c)$ relations (\ref{7}) are convergent at $\eta=1$. For
more analysis, the following vectors are defined as
$$
\begin{array}{l}
\displaystyle  \tilde{\mathcal{T}}_N(c) = \bigg \{ \mathcal{T}_0(c), \mathcal{T}_1(c), \ldots, \mathcal{T}_N(c) \bigg\},  \\
\\
\displaystyle  \tilde{\mathcal{I}}_N(c) = \bigg \{ \mathcal{I}_0(c), \mathcal{I}_1(c), \ldots, \mathcal{I}_N(c) \bigg\},  \\
  \\
\displaystyle    \tilde{\mathcal{V}}_N(c) = \bigg \{ \mathcal{V}_0(c), \mathcal{V}_1(c), \ldots, \mathcal{V}_N(c) \bigg\}. \\
\end{array}
$$
Differentiating the Eqs. (\ref{6}) $d$-times with respect to $\eta$,
dividing by $d!$ and putting $\eta=0$ the $d$-th order deformation
equations for vectors $\vec{\mathcal{T}}, \vec{\mathcal{I}},
\vec{\mathcal{V}}$ can be obtained
\begin{equation}\label{8}
\begin{array}{l}
\displaystyle L_1 \left[\mathcal{T}_d(c) - \chi_d \mathcal{T}_{d-1}(c)\right] = \hbar H_1 (c) \Re_d^1 \left(\vec{\mathcal{T}}_{d-1}, \vec{\mathcal{I}}_{d-1}, \vec{\mathcal{V}}_{d-1}\right),\\
\\
\displaystyle L_2 \left[\mathcal{I}_d(c) - \chi_d \mathcal{I}_{d-1}(c)\right] = \hbar H_2 (c) \Re_d^2 \left(\vec{\mathcal{T}}_{d-1}, \vec{\mathcal{I}}_{d-1}, \vec{\mathcal{V}}_{d-1}\right),\\
 \\
\displaystyle  L_3 \left[\mathcal{V}_d(c) - \chi_d \mathcal{V}_{d-1}(c)\right] = \hbar H_3 (c) \Re_d^3 \left(\vec{\mathcal{T}}_{d-1}, \vec{\mathcal{I}}_{d-1}, \vec{\mathcal{V}}_{d-1}\right),\\
\end{array}
\end{equation}
where
$$
\mathcal{T}_d(0) = 0,~~~~~ \mathcal{I}_d(0) = 0,~~~~~
\mathcal{V}_d(0) = 0,
$$
and
\begin{equation}\label{9}
\begin{array}{ll}
\displaystyle \Re_d^1 (c)& \displaystyle = \mathcal{L}
\bigg[\mathcal{T}_{d-1}(c)\bigg]
 - \frac{\mathcal{T}_{d-1}(0)}{z} -  (1- \chi_d)\frac{\mathcal{L} \bigg[s\bigg]}{z}
- \frac{(g-\omega_\mathcal{T})}{z} \mathcal{L}
\bigg[\mathcal{T}_{d-1}(c)\bigg]
  \\
 \\  &\displaystyle + \frac{g}{z \mathcal{T}_{\max}}  \mathcal{L} \bigg[ \sum_{p=0}^{d-1}\mathcal{T}_p(c) \mathcal{T}_{d-1-p}(c) \bigg] + \frac{g}{z \mathcal{T}_{\max}} \mathcal{L} \bigg[\sum_{p=0}^{d-1}\mathcal{T}_p(c) \mathcal{I}_{d-1-p}(c)\bigg] \\
 \\ & \displaystyle  + \frac{\alpha}{z} \mathcal{L} \bigg[\sum_{p=0}^{d-1}\mathcal{V}_p(c) \mathcal{T}_{d-1-p}(c)\bigg], \\
   \\
\displaystyle \Re_d^2 (c) &\displaystyle  = \mathcal{L} \bigg[\mathcal{I}_{d-1}(c)\bigg]  \displaystyle - \frac{\mathcal{I}_{d-1}(0)}{z} -  \frac{\beta}{z} \mathcal{L} \bigg[\sum_{p=0}^{d-1} \mathcal{V}_p(c) \mathcal{T}_{d-1-p}(c) \bigg]  \\
\\
& \displaystyle + \frac{\omega_\mathcal{I}}{z} \mathcal{L} \bigg[\mathcal{I}_{d-1}(c)\bigg],\\
 \\
\displaystyle \Re_d^3 (c) & \displaystyle  = \mathcal{L}
\bigg[\mathcal{V}_{d-1}(c)\bigg] \displaystyle  -
\frac{\mathcal{V}_{d-1}(0)}{z} - \frac{n \omega_b}{z} \mathcal{L}
\bigg[\mathcal{I}_{d-1}(c)\bigg] \\
\\ & \displaystyle + \frac{\alpha}{z}
\mathcal{L}\bigg[ \sum_{p=0}^{d-1}\mathcal{V}_p(c)
\mathcal{T}_{d-1-p}(c)\bigg] + \frac{\omega_\mathcal{V}}{z}
\mathcal{L}\bigg[\mathcal{V}_{d-1}(c)\bigg],
\end{array}
\end{equation}
and
$$
\chi_d = \left\{ \begin{array}{l}
                   0, ~~~~~d\leq 1 \\
                     \\
                   1, ~~~~~d > 1.
                 \end{array}\right.
$$

Using $H_1(c) = H_2(c) = H_3(c) = 1$ and applying the inverse
operator  $L^{-1} = \mathcal{L}^{-1}$ the solutions of $d$-th order
deformation equations (\ref{8}) are obtained as
$$
\begin{array}{l}
\mathcal{T}_d(c) - \chi_d \mathcal{T}_{d-1}(c) \\
\\
\displaystyle = \hbar \bigg\{ \mathcal{T}_{d-1}(c)
  -  (1- \chi_d) \mathcal{L}^{-1} \bigg[ \frac{\mathcal{L} \bigg[s\bigg]}{z}\bigg]
- \mathcal{L}^{-1} \bigg[ \frac{(g-\omega_\mathcal{T})}{z}
\mathcal{L} \bigg[\mathcal{T}_{d-1}(c)\bigg] \bigg]
  \\
 \\   \displaystyle + \mathcal{L}^{-1} \bigg[ \frac{g}{z \mathcal{T}_{\max}}  \mathcal{L} \bigg[ \sum_{p=0}^{d-1}\mathcal{T}_p(c) \mathcal{T}_{d-1-p}(c) \bigg] \bigg]  \\
 \\   \displaystyle  + \mathcal{L}^{-1} \bigg[ \frac{g}{z \mathcal{T}_{\max}} \mathcal{L} \bigg[\sum_{p=0}^{d-1}\mathcal{T}_p(c) \mathcal{I}_{d-1-p}(c)\bigg]\bigg] + \mathcal{L}^{-1} \bigg[ \frac{\alpha}{z} \mathcal{L} \bigg[\sum_{p=0}^{d-1}\mathcal{V}_p(c) \mathcal{T}_{d-1-p}(c)\bigg]\bigg]    \bigg \}    ,\\
\end{array}
$$
\begin{equation}\label{10}
\begin{array}{l}
 \displaystyle \mathcal{I}_d(c) - \chi_d \mathcal{I}_{d-1}(c) \\
 \\
  \displaystyle = \hbar \bigg\{  \mathcal{I}_{d-1}(c)  -  \mathcal{L}^{-1} \bigg[ \frac{\beta}{z} \mathcal{L} \bigg[\sum_{p=0}^{d-1} \mathcal{V}_p(c) \mathcal{T}_{d-1-p}(c) \bigg] \bigg] + \mathcal{L}^{-1} \bigg[ \frac{\omega_\mathcal{I}}{z} \mathcal{L} \bigg[\mathcal{I}_{d-1}(c)\bigg]\bigg] \bigg \}    ,\\
 \\
 \displaystyle \mathcal{V}_d(c) - \chi_d \mathcal{V}_{d-1}(c)\\
 \\
   \displaystyle = \hbar  \bigg\{  \mathcal{V}_{d-1}(c) - \mathcal{L}^{-1} \bigg[ \frac{n
\omega_b}{z} \mathcal{L} \bigg[\mathcal{I}_{d-1}(c)\bigg]\bigg] +
\mathcal{L}^{-1} \bigg[ \frac{\alpha}{z}
\mathcal{L}\bigg[ \sum_{p=0}^{d-1}\mathcal{V}_p(c) \mathcal{T}_{d-1-p}(c)\bigg]\bigg]\\
\\  \displaystyle + \mathcal{L}^{-1} \bigg[ \frac{\omega_\mathcal{V}}{z} \mathcal{L}\bigg[\mathcal{V}_{d-1}(c)\bigg] \bigg]  \bigg \}.\\
\end{array}
\end{equation}
Finally, the $N$-th order of approximate solutions can be obtained
by
\begin{equation}\label{11}
\mathcal{T}_N(c) = \sum_{p=0}^{N} \mathcal{T}_p(c),
~~~~~\mathcal{I}_N(c) = \sum_{p=0}^{N} \mathcal{I}_p(c), ~~~~~
\mathcal{V}_N(c) = \sum_{p=0}^{N} \mathcal{V}_p(c).
\end{equation}
\section{Convergence theorem}
In this section, by presenting a theorem, convergence of HATM to
solve the non-linear system of equations (\ref{1}) is illustrated.

\begin{theorem} If series solutions
\begin{equation}\label{12}
\begin{array}{l}
\displaystyle \mathcal{T}(c) = \mathcal{T}_0(c) +
\sum_{d=1}^{\infty} \mathcal{T}_d(c),
\\
\\ \displaystyle \mathcal{I}(c) = \mathcal{I}_0(c) +
\sum_{d=1}^{\infty} \mathcal{I}_d(c),
\\
\\ \displaystyle \mathcal{V}(c) = \mathcal{V}_0(c) +
\sum_{d=1}^{\infty} \mathcal{V}_d(c),
\end{array}
\end{equation}
are convergent where $\mathcal{T}_d(c), \mathcal{I}_d(c),
\mathcal{V}_d(c)$ are formed by Eqs. (\ref{8}), then they must be
the exact solution of system (\ref{1}).

\textbf{Proof:} Assume that the series solutions (\ref{12}) are
convergent. Therefore we get
\begin{equation}\label{13}
\begin{array}{lll}
\displaystyle S_1(c) = \sum_{d=0}^{\infty} \mathcal{T}_d(c),~~~~~ &
\displaystyle S_2(c) =  \sum_{d=0}^{\infty} \mathcal{I}_d(c),~~~~~ &
\displaystyle S_3(c) = \sum_{d=0}^{\infty} \mathcal{V}_d(c),
\end{array}
\end{equation}
where we have
\begin{equation}\label{14}
\begin{array}{lll}
\lim_{d \rightarrow \infty} \mathcal{T}_d(c) = 0,~~~~~& \lim_{d
\rightarrow \infty} \mathcal{I}_d(c) = 0,~~~~~& \lim_{d \rightarrow
\infty} \mathcal{V}_d(c) = 0.
\end{array}
\end{equation}
By mentioned details of Section 2 we can write
\begin{equation}\label{15}
\begin{array}{l}
\displaystyle  \sum_{d=1}^N \bigg[\mathcal{T}_d(c) - \chi_d \mathcal{T}_{d-1}(c)\bigg] =  \mathcal{T}_N(c),\\
  \\
\displaystyle  \sum_{d=1}^N \bigg[\mathcal{I}_d(c) - \chi_d \mathcal{I}_{d-1}(c)\bigg] =  \mathcal{I}_N(c),\\
  \\
\displaystyle  \sum_{d=1}^N \bigg[\mathcal{V}_d(c) - \chi_d
\mathcal{V}_{d-1}(c)\bigg] = \mathcal{V}_N(c),
\end{array}
\end{equation}
which by using Eqs. (\ref{14}) and (\ref{15}) the following
relations can be obtained
\begin{equation}\label{16}
\begin{array}{l}
\displaystyle  \sum_{d=1}^N \bigg[\mathcal{T}_d(c) - \chi_d \mathcal{T}_{d-1}(c)\bigg] = \lim_{N \rightarrow \infty} \mathcal{T}_N(c) = 0,\\
  \\
\displaystyle  \sum_{d=1}^N \bigg[\mathcal{I}_d(c) - \chi_d
\mathcal{I}_{d-1}(c)\bigg] = \lim_{N \rightarrow \infty}
\mathcal{I}_N(c) = 0,\\
  \\
\displaystyle  \sum_{d=1}^N \bigg[\mathcal{V}_d(c) - \chi_d
\mathcal{V}_{d-1}(c)\bigg] = \lim_{N \rightarrow \infty}
\mathcal{V}_N(c) = 0.
\end{array}
\end{equation}
Since the operator $L$ is a linear operator thus
\begin{equation}\label{17}
\begin{array}{l}
\displaystyle  \sum_{d=1}^\infty L \bigg[\mathcal{T}_d(c) - \chi_d \mathcal{T}_{d-1}(c)\bigg] = L \bigg[ \sum_{d=1}^\infty \mathcal{T}_d(c) - \chi_d \mathcal{T}_{d-1}(c)\bigg]= 0 ,\\
  \\
\displaystyle  \sum_{d=1}^\infty L \bigg[\mathcal{I}_d(c) - \chi_d \mathcal{I}_{d-1}(c)\bigg] =  L \bigg[ \sum_{d=1}^\infty \mathcal{I}_d(c) - \chi_d \mathcal{I}_{d-1}(c)\bigg]= 0,\\
  \\
\displaystyle  \sum_{d=1}^\infty L \bigg[\mathcal{V}_d(c) - \chi_d
\mathcal{V}_{d-1}(c)\bigg] =  L \bigg[ \sum_{d=1}^\infty
\mathcal{V}_d(c) - \chi_d \mathcal{V}_{d-1}(c)\bigg]= 0,
\end{array}
\end{equation}
and we can obtain
\begin{equation}\label{18}
\begin{array}{l}
\displaystyle  \sum_{d=1}^\infty L_1 \bigg[\mathcal{T}_d(c) - \chi_d \mathcal{T}_{d-1}(c)\bigg] = \hbar H_1(c)  \sum_{d=1}^\infty \Re_d^1 (\vec{\mathcal{T}}_{d-1}, \vec{\mathcal{I}}_{d-1}, \vec{\mathcal{V}}_{d-1} ) =0,\\
  \\
\displaystyle  \sum_{d=1}^\infty L_2 \bigg[\mathcal{I}_d(c) - \chi_d \mathcal{I}_{d-1}(c)\bigg] = \hbar H_2(c)  \sum_{d=1}^\infty \Re_d^2 (\vec{\mathcal{T}}_{d-1}, \vec{\mathcal{I}}_{d-1}, \vec{\mathcal{V}}_{d-1} ) =0,\\
  \\
\displaystyle  \sum_{d=1}^\infty L_3 \bigg[\mathcal{V}_d(c) - \chi_d
\mathcal{V}_{d-1}(c)\bigg] = \hbar H_3(c)  \sum_{d=1}^\infty \Re_d^3
(\vec{\mathcal{T}}_{d-1}, \vec{\mathcal{I}}_{d-1},
\vec{\mathcal{V}}_{d-1} ) =0.
\end{array}
\end{equation}
Since in Eqs. (\ref{18}), $\hbar \neq 0, H_1(c)\neq 0, H_2(c)\neq 0,
H_3(c)\neq 0$ we get
\begin{equation}\label{19}
\begin{array}{l}
\displaystyle    \sum_{d=1}^\infty \Re_d^1 (\vec{\mathcal{T}}_{d-1}, \vec{\mathcal{I}}_{d-1}, \vec{\mathcal{V}}_{d-1} ) =0,\\
  \\
\displaystyle   \sum_{d=1}^\infty \Re_d^2 (\vec{\mathcal{T}}_{d-1}, \vec{\mathcal{I}}_{d-1}, \vec{\mathcal{V}}_{d-1} ) =0,\\
  \\
\displaystyle   \sum_{d=1}^\infty \Re_d^3 (\vec{\mathcal{T}}_{d-1}, \vec{\mathcal{I}}_{d-1}, \vec{\mathcal{V}}_{d-1} ) =0.\\
\end{array}
\end{equation}
By putting $\Re_d^1 (\mathcal{T}_{d-1}(c) ), \Re_d^2
(\mathcal{I}_{d-1}(c))$ and $\Re_d^3 (\mathcal{V}_{d-1}(c))$ into
Eqs. (\ref{19}) we get
\begin{equation}\label{20}
\begin{array}{l}
\displaystyle    \sum_{d=1}^\infty \Re_d^1 (\mathcal{T}_{d-1}(c) ) \\
\\
\displaystyle = \sum_{d=1}^\infty    \bigg[ \mathcal{T}'_{d-1}(c)
  -  (1- \chi_d)s
-  (g-\omega_\mathcal{T})   \mathcal{T}_{d-1}(c)
   \displaystyle + \frac{g}{  \mathcal{T}_{\max}}   \sum_{p=0}^{d-1}\mathcal{T}_p(c)
   \mathcal{T}_{d-1-p}(c)\\
   \\
   ~~~~~~~~~~ \displaystyle + \frac{g}{ \mathcal{T}_{\max}}
\sum_{p=0}^{d-1}\mathcal{T}_p(c)
   \mathcal{I}_{d-1-p}(c) +  \alpha  \sum_{p=0}^{d-1}\mathcal{V}_p(c) \mathcal{T}_{d-1-p}(c)\bigg] \\
   \\
   \displaystyle =    \sum_{d=0}^\infty  \mathcal{T}'_{d}(c)
  -   s
-  (g-\omega_\mathcal{T})  \sum_{d=0}^\infty  \mathcal{T}_{d}(c)
    + \frac{g}{  \mathcal{T}_{\max}}   \sum_{d=1}^\infty  \sum_{p=0}^{d-1}\mathcal{T}_p(c)
   \mathcal{T}_{d-1-p}(c)\\
   \\
   ~~~~~~~~~~ \displaystyle + \frac{g}{ \mathcal{T}_{\max}}
\sum_{d=1}^\infty \sum_{p=0}^{d-1}\mathcal{T}_p(c)
   \mathcal{I}_{d-1-p}(c) + \alpha   \sum_{d=1}^\infty  \sum_{p=0}^{d-1}\mathcal{V}_p(c)
   \mathcal{T}_{d-1-p}(c)\\
   \\
   \displaystyle =   \sum_{d=0}^\infty  \mathcal{T}'_{d}(c)
  -   s
-  (g-\omega_\mathcal{T})  \sum_{d=0}^\infty  \mathcal{T}_{d}(c)
    + \frac{g}{  \mathcal{T}_{\max}}   \sum_{p=0}^\infty  \sum_{d=p+1}^{\infty}\mathcal{T}_p(c)
   \mathcal{T}_{d-1-p}(c)\\
   \\
   ~~~~~~~~~~ \displaystyle + \frac{g}{ \mathcal{T}_{\max}}
\sum_{p=0}^\infty \sum_{d=p+1}^{\infty} \mathcal{T}_p(c)
   \mathcal{I}_{d-1-p}(c) + \alpha   \sum_{p=0}^\infty  \sum_{d=p+1}^{\infty} \mathcal{V}_p(c)
   \mathcal{T}_{d-1-p}(c)\\
   \\
    \displaystyle =   \sum_{d=0}^\infty  \mathcal{T}'_{d}(c)
  -   s
-  (g-\omega_\mathcal{T})  \sum_{d=0}^\infty  \mathcal{T}_{d}(c)
    + \frac{g}{ \mathcal{T}_{\max}}   \sum_{p=0}^\infty \mathcal{T}_p(c) \sum_{d=0}^{\infty}
   \mathcal{T}_{d}(c)\\
   \\
   ~~~~~~~~~~ \displaystyle + \frac{g}{ \mathcal{T}_{\max}}
\sum_{p=0}^\infty \mathcal{T}_p(c) \sum_{d=0}^{\infty}
   \mathcal{I}_{d}(c) + \alpha   \sum_{p=0}^\infty \mathcal{V}_p(c) \sum_{d=0}^{\infty}
   \mathcal{T}_{d}(c)\\
   \\
\displaystyle   = S'_1 (c) - s - (g-\omega_\mathcal{T}) S_1(c) +
\frac{g}{ \mathcal{T}_{\max}} S_1^2 (c)\\
\\
 ~~~~~~~~~~ \displaystyle
   + \frac{g}{ \mathcal{T}_{\max}} S_1(c) S_2(c) + \alpha   S_3(c) S_1(c)=0,
\end{array}
\end{equation}
and
\begin{equation}\label{21}
 \begin{array}{ll}
\displaystyle \sum_{d=1}^\infty  \Re_d^2 (\mathcal{I}_{d-1}(c)) &\displaystyle = \sum_{d=1}^\infty \bigg[\mathcal{I}'_{d-1}(c) -  \beta \sum_{p=0}^{d-1} \mathcal{V}_p(c) \mathcal{T}_{d-1-p}(c)   + \omega_\mathcal{I} \mathcal{I}_{d-1}(c)\bigg]\\
\\
& \displaystyle  = \sum_{d=0}^\infty \mathcal{I}'_{d}(c) -  \beta \sum_{d=1}^\infty \sum_{p=0}^{d-1} \mathcal{V}_p(c) \mathcal{T}_{d-1-p}(c) + \omega_\mathcal{I} \sum_{d=0}^\infty \mathcal{I}_{d}(c)\\
\\
& \displaystyle  = \sum_{d=0}^\infty \mathcal{I}'_{d}(c) -  \beta \sum_{p=0}^\infty \sum_{d=p+1}^{\infty} \mathcal{V}_p(c) \mathcal{T}_{d-1-p}(c)  + \omega_\mathcal{I} \sum_{d=0}^\infty \mathcal{I}_{d}(c)\\
\\
& \displaystyle  = \sum_{d=0}^\infty \mathcal{I}'_{d}(c) -  \beta \sum_{p=0}^\infty \mathcal{V}_p(c) \sum_{d=0}^{\infty}  \mathcal{T}_{d}(c)  + \omega_\mathcal{I} \sum_{d=0}^\infty \mathcal{I}_{d}(c)\\
\\
& \displaystyle  = S'_2(c) - \beta S_3(c) S_1(c) +
\omega_\mathcal{I} S_2(c)=0,
 \end{array}
\end{equation}
and finally
  \begin{equation}\label{22}
\begin{array}{l}
 \displaystyle \sum_{d=1}^\infty \Re_d^3 (\mathcal{V}_{d-1}(c)) \\
 \\
  \displaystyle = \sum_{d=1}^\infty \bigg[ \mathcal{V}'_{d-1}(c) - n \omega_b
\mathcal{I}_{d-1}(c)+ \alpha \sum_{p=0}^{d-1}\mathcal{V}_p(c)
\mathcal{T}_{d-1-p}(c)
  + \omega_\mathcal{V} \mathcal{V}_{d-1}(c) \bigg]\\
  \\
  \displaystyle = \sum_{d=0}^\infty \mathcal{V}'_{d}(c) - n \omega_b
\sum_{d=0}^\infty \mathcal{I}_{d}(c)+ \alpha \sum_{d=1}^\infty
\sum_{p=0}^{d-1}\mathcal{V}_p(c) \mathcal{T}_{d-1-p}(c)
  + \omega_\mathcal{V} \sum_{d=0}^\infty \mathcal{V}_{d}(c) \\
  \\
    \displaystyle = \sum_{d=0}^\infty \mathcal{V}'_{d}(c) - n \omega_b
\sum_{d=0}^\infty \mathcal{I}_{d}(c)+ \alpha \sum_{p=0}^\infty
\sum_{d=p+1}^{\infty}\mathcal{V}_p(c) \mathcal{T}_{d-1-p}(c)
  + \omega_\mathcal{V} \sum_{d=0}^\infty \mathcal{V}_{d}(c)\\
\\
 \displaystyle = \sum_{d=0}^\infty \mathcal{V}'_{d}(c) - n \omega_b
\sum_{d=0}^\infty \mathcal{I}_{d}(c)+ \alpha \sum_{p=0}^\infty
\mathcal{V}_p(c) \sum_{d=0}^{\infty} \mathcal{T}_{d}(c)
  + \omega_\mathcal{V} \sum_{d=0}^\infty \mathcal{V}_{d}(c) \\
  \\
 \displaystyle = S'_3 (c) - n \omega_b S_2(c) + \alpha S_3(c) S_2(c) + \omega_\mathcal{V}  S_3(c)=0.
 \end{array}
\end{equation}
Eqs. (\ref{20}), (\ref{21}) and (\ref{22}) show that the series
solutions $S_1(c), S_2(c)$ and $S_3(c)$ must be the exact solutions
of system of equations (\ref{1}).
\end{theorem}
\section{Numerical investigations}
\begin{figure}
\centering
$$\begin{array}{c}
 \includegraphics[width=3in]{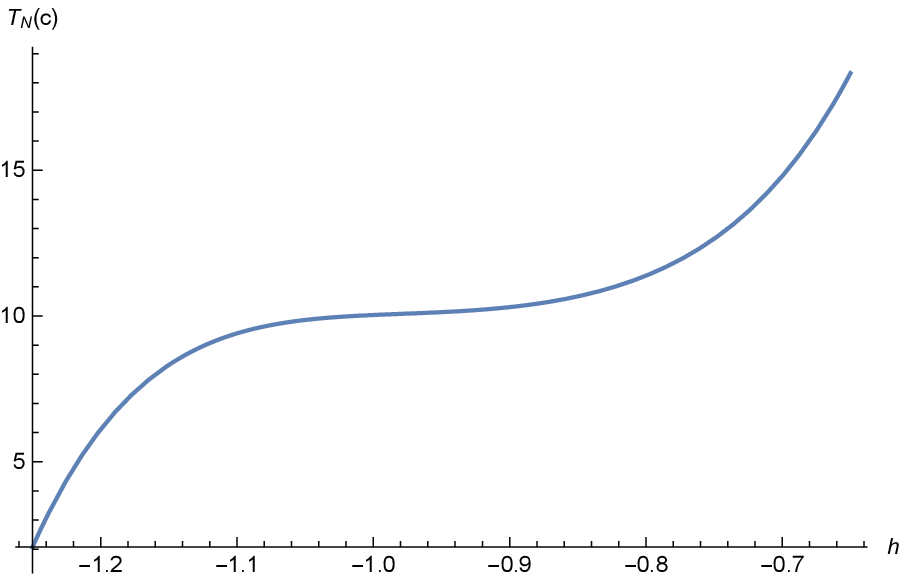}\\
 \\
 \\
 \\
 \includegraphics[width=3in]{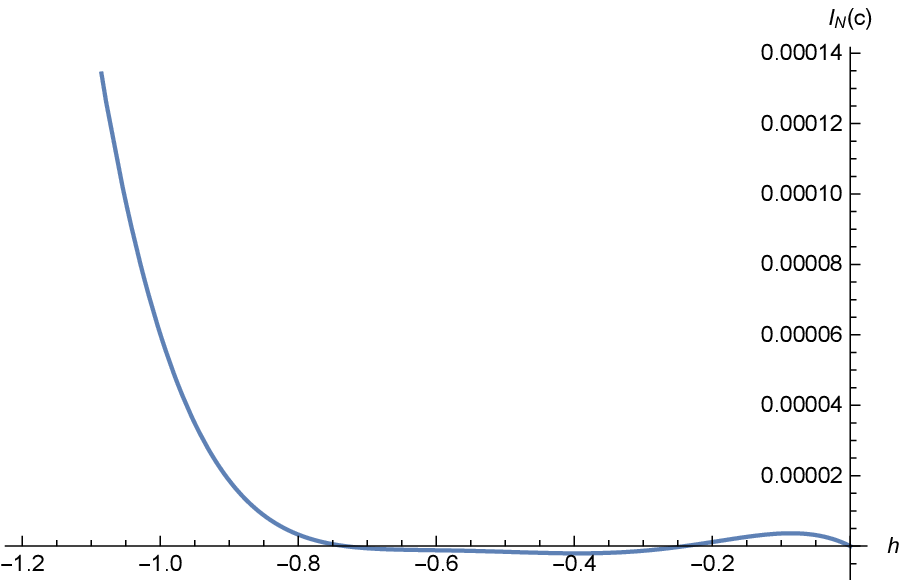}\\
 \\
 \\
 \\
 \includegraphics[width=3in]{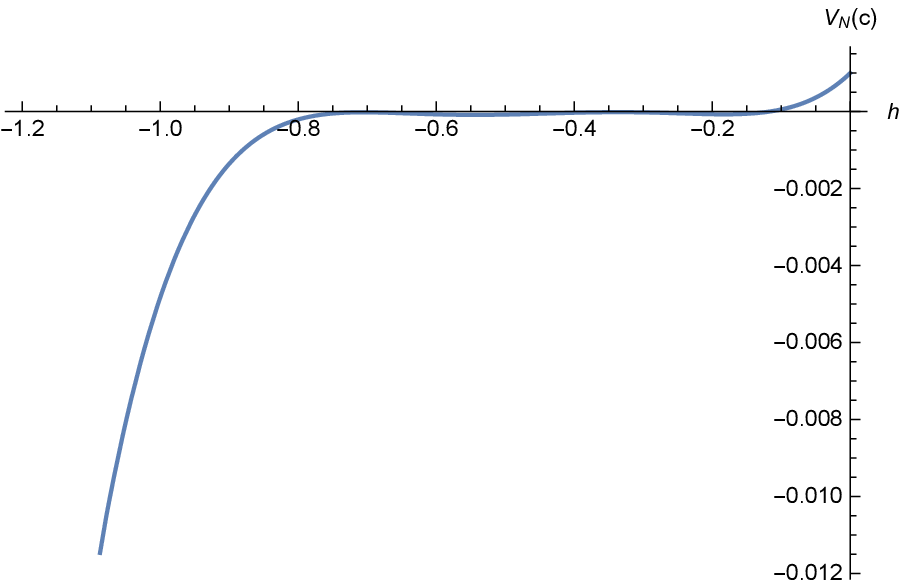}
\end{array}$$
  \caption{$\hbar$-curves of $\mathcal{T}_N(c), \mathcal{I}_N(c)$ and $\mathcal{V}_N(c)$ for $N=5, c=1$. }\label{f1}
\end{figure}

\begin{figure}
\centering
$$\begin{array}{c}
 \includegraphics[width=3in]{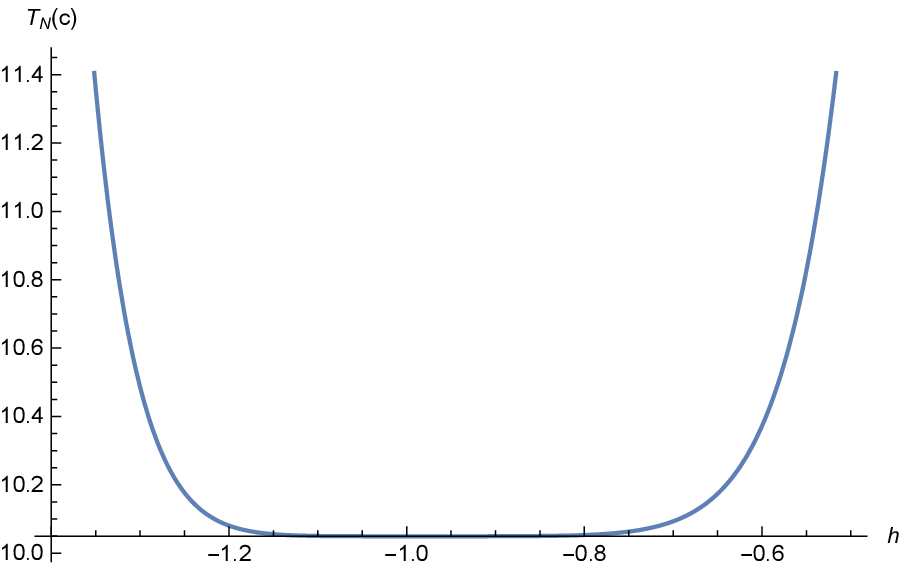}\\
 \\
 \\
 \\
 \includegraphics[width=3in]{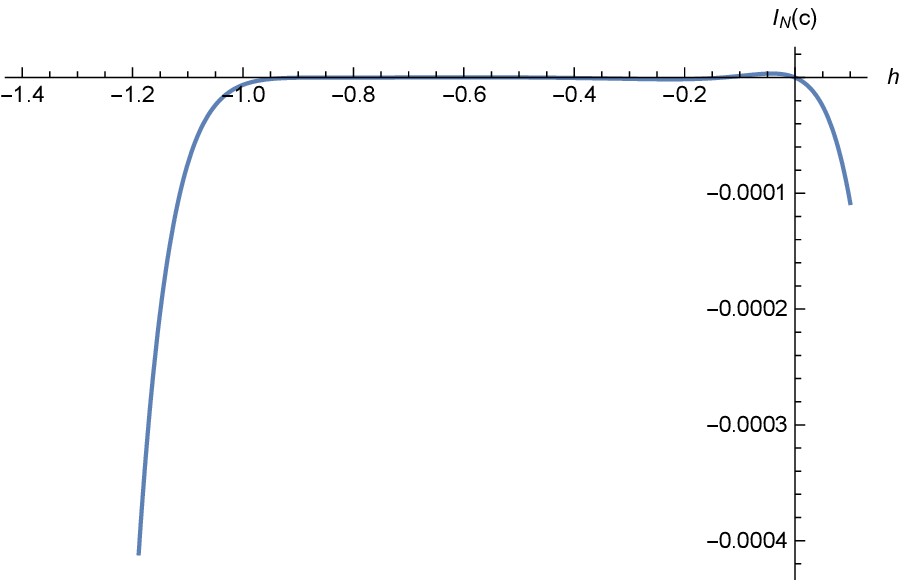}\\
 \\
 \\
 \\
 \includegraphics[width=3in]{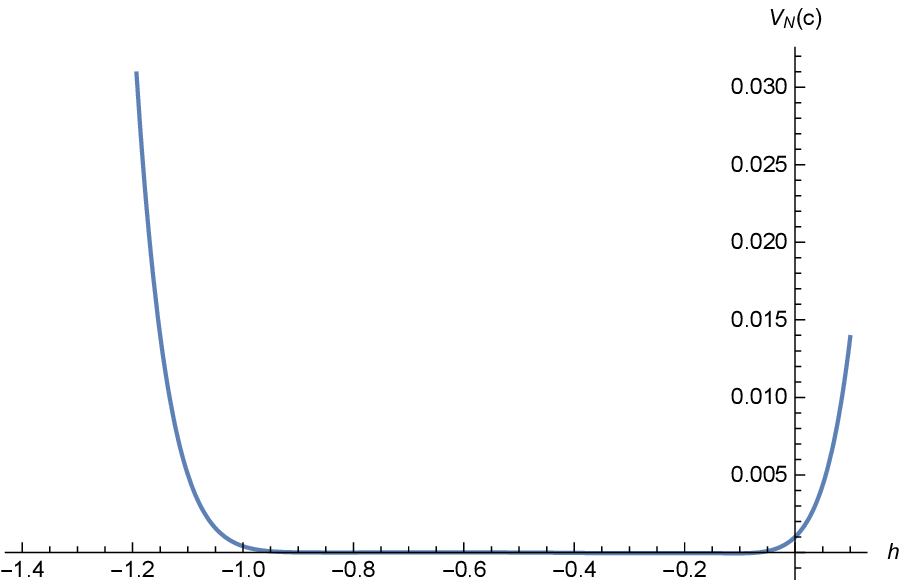}
\end{array}$$
  \caption{$\hbar$-curves of $\mathcal{T}_N(c), \mathcal{I}_N(c)$ and $\mathcal{V}_N(c)$ for $N=10, c=1$. }\label{f2}
\end{figure}

\begin{figure}
\centering
$$\begin{array}{c}
 \includegraphics[width=3in]{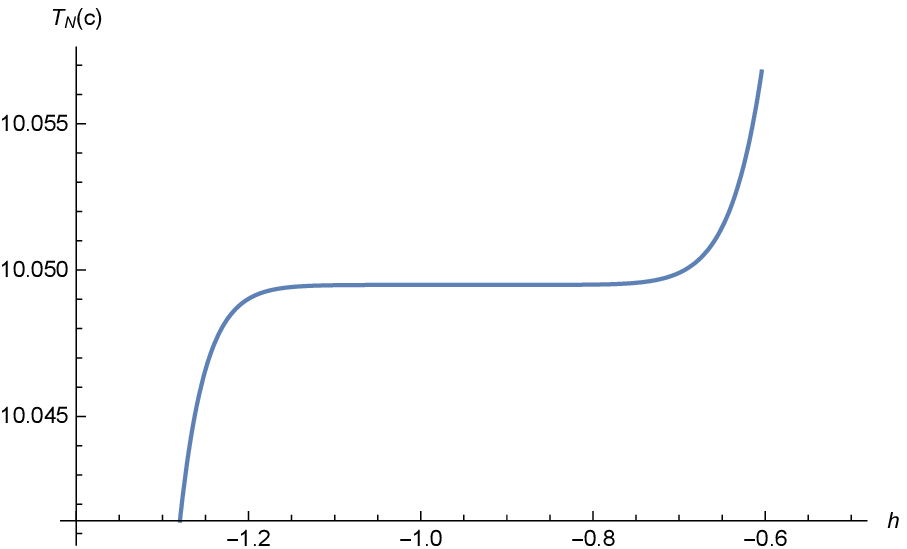}\\
 \\
 \\
 \\
 \includegraphics[width=3in]{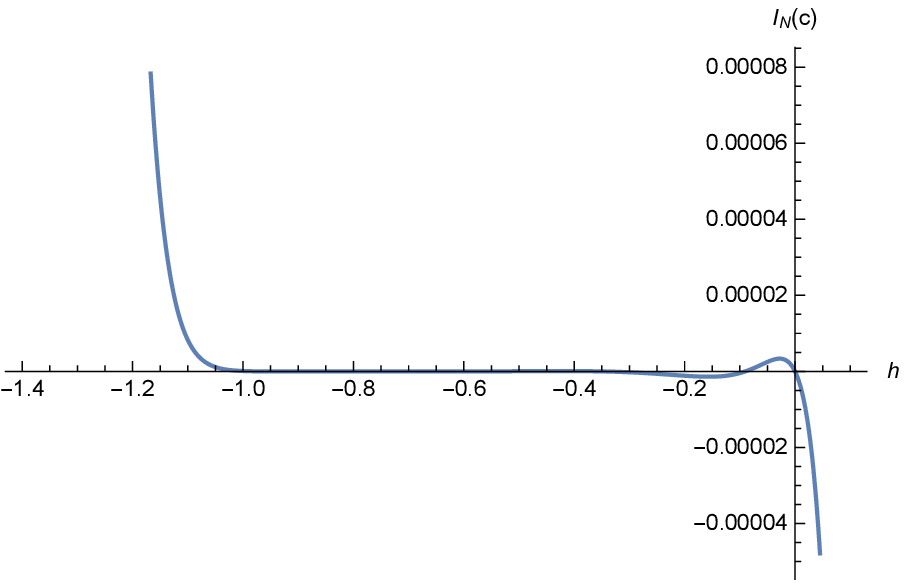}\\
 \\
 \\
 \\
 \includegraphics[width=3in]{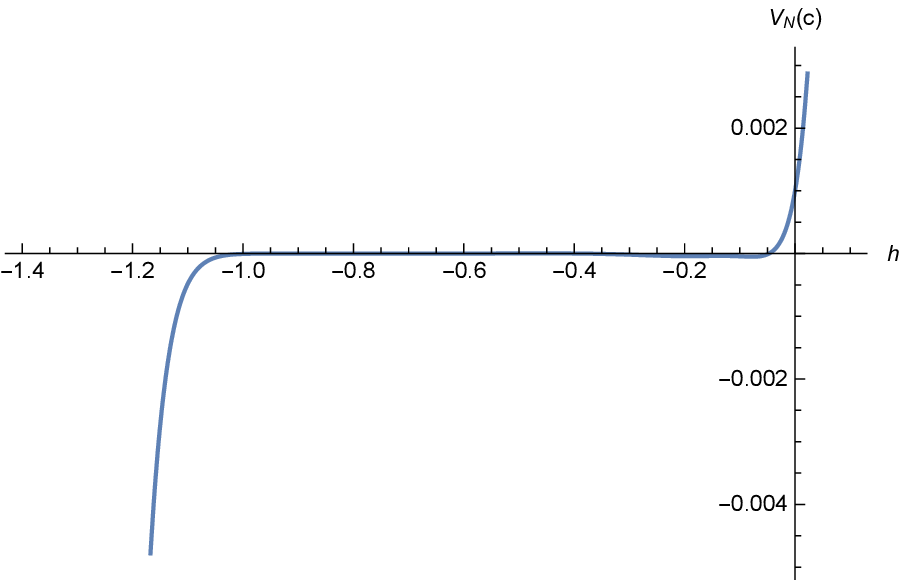}
\end{array}$$
  \caption{$\hbar$-curves of $\mathcal{T}_N(c), \mathcal{I}_N(c)$ and $\mathcal{V}_N(c)$ for $N=15, c=1$. }\label{f3}
\end{figure}

In this section, the HATM is applied to solve the non-linear system
of Eqs. (\ref{1}) for mentioned values of Table \ref{t0}. Several
$\hbar$-curves are demonstrated in Figs. \ref{f1}, \ref{f2} and
\ref{f3} for various $N$ to show the regions of convergence. We note
that these regions are parallel parts with axiom $x$. According to
these figures, the convergence regions for $N=5,10,15$ are presented
in table \ref{tt}.

\begin{table}[h]
\caption{ Regions of convergence for $N=5, 10, 15$ and $c=1$.
}\label{tt}
 \centering
\begin{tabular}{|c|c|c|c|}
  \hline
  $N$ & $\hbar_{\mathcal{T}}$ &$\hbar_{\mathcal{I}}$&$\hbar_{\mathcal{V}}$  \\
  \hline
  5&$ -1.1\leq \hbar_{\mathcal{T}} \leq -0.9$&$-1.1\leq \hbar_{\mathcal{I}} \leq -0.8 $&$ -1.2 \leq \hbar_{\mathcal{V}} \leq -0.7 $\\
  &&&\\
 10 &$ -0.8 \leq \hbar_{\mathcal{T}} \leq -0.2$&$-1 \leq \hbar_{\mathcal{I}} \leq -0.2$&$ -1 \leq \hbar_{\mathcal{V}} \leq -0.2$\\
 &&&\\
  15 &$ -0.8 \leq \hbar_{\mathcal{T}} \leq -0.2$&$-1 \leq \hbar_{\mathcal{I}} \leq -0.1$&$ -1 \leq \hbar_{\mathcal{V}} \leq -0.1 $\\
  \hline
\end{tabular}
\end{table}

The approximate solution of problem (\ref{1}) for $N=5$ is in the
following form
$$\left\{
\begin{array}{l}
  \mathcal{T}_{5}(c) =1000 + 5000 \hbar + 10000 \hbar^2  +\cdots +
 0.000017219 \hbar^5 c^5,\\
\\
 \mathcal{I}_{5}(c) = - 0.0001 \hbar c - 0.0006 \hbar^2 c + \cdots -
  0.0000155458 \hbar^5 c^5,\\
\\
 \mathcal{V}_{5}(c) = 0.001 + 0.005 \hbar + 0.01 \hbar^2 + \cdots +
  0.00230941 \hbar^5 c^5,
\end{array}\right.
$$
and for $N=10$ we get
$$\left\{
\begin{array}{l}
  \mathcal{T}_{10}(c) = 1000 + 10000 \hbar + 45000 \hbar^2 + \cdots  + 2.00653\times 10^{-7} \hbar^{10} c^{10}, \\
    \\
 \mathcal{I}_{10}(c) =
 - 0.0002 \hbar c- 0.0027\hbar^2 c +
 \cdots - 1.81169\times 10^{-7}
 \hbar^{10}
 c^{10},\\
 \\
 \mathcal{V}_{10}(c) = 0.001 + 0.01 \hbar + 0.045 \hbar^2  +
 \cdots + 0.00002 \hbar^{10} c^{10}.
\end{array}\right.
$$
Finally, for $N=15$ the approximate solution is obtained as follows
$$\left\{
\begin{array}{l}
  \mathcal{T}_{15}(c) = 1000 + 15000 \hbar + 105000  \hbar^2 +\cdots    + 1.96276\times 10^{-10} \hbar^{15} c^{15},\\
  \\
   \mathcal{I}_{15}(c) = - 0.0003 \hbar c - 0.0063 \hbar^2 c +\cdots  - 1.77216\times 10^{-10} \hbar^{15} c^{15},\\
   \\
    \mathcal{V}_{15}(c) = 0.001 + 0.015 \hbar + 0.105
    \hbar^2 +\cdots  + 2.63289\times 10^{-8} \hbar^{15}
    c^{15}.
\end{array}\right.
$$

In order to show the efficiency and accuracy of presented method the
following residual errors
\begin{equation}\label{4-1}
\begin{array}{l}
\displaystyle  R_1(\mathcal{T},\mathcal{I},\mathcal{V};\hbar)  =\frac{d \mathcal{T}_N(c;\hbar)}{dt} - s + \omega_\mathcal{T} \mathcal{T}_N(c;\hbar) \\
\\   ~~~~~~~~~~~~~~~~~~~~ \displaystyle - r \mathcal{T}_N(c;\hbar) \left(1-\frac{\mathcal{I}_N(c;\hbar)+\mathcal{T}_N(c;\hbar)}{\mathcal{T}_{\max}}\right) \\
\\
 ~~~~~~~~~~~~~~~~~~~~ \displaystyle + \alpha \mathcal{V}_N(c;\hbar) \mathcal{T}_N(c;\hbar), \\
   \\
\displaystyle R_2(\mathcal{T},\mathcal{I},\mathcal{V};\hbar)= \frac{d\mathcal{I}_N(c;\hbar)}{dt} - \beta \mathcal{V}_N(c;\hbar) \mathcal{T}_N(c;\hbar) + \omega_\mathcal{I} \mathcal{I}_N(c;\hbar),\\
 \\
\displaystyle R_3(\mathcal{T},\mathcal{I},\mathcal{V};\hbar)=
\frac{d\mathcal{V}_N(c;\hbar)}{dt} - n \omega_b
\mathcal{I}_N(c;\hbar) + \alpha \mathcal{V}_N(c;\hbar)
\mathcal{T}_N(c;\hbar) \\
\\
~~~~~~~~~~~~~~~~~~~~ \displaystyle + \omega_\mathcal{V}
\mathcal{V}_N(c;\hbar),
\end{array}
\end{equation}
are applied. In Tables \ref{t1} and \ref{t2} the errors of the
residual functions (\ref{4-1}) are shown for different values of $c$
and $N=10, 15$. Also, the errors $R_1, R_2, R_3$ for various
$\hbar,c$ and $N=5,10,15$ are presented in Tables \ref{t3}, \ref{t4}
and \ref{t5}.

\begin{table}
\caption{ The residual errors $R_1, R_2, R_3$ for different values
of $c$ and $N=10$. }\label{t1}
 \centering
\scalebox{0.9}{
\begin{tabular}{|c|c|c|c|}
  \hline
$c$ & $R_1(\mathcal{T}_{10}, \mathcal{I}_{10}, \mathcal{V}_{10}; \hbar = -1)$  &  $R_2(\mathcal{T}_{10}, \mathcal{I}_{10}, \mathcal{V}_{10}; \hbar = -0.8)$   &  $R_3(\mathcal{T}_{10}, \mathcal{I}_{10}, \mathcal{V}_{10}; \hbar = -0.8)$   \\
    \hline
$0.0$&$ 0 $             &$1.39264\times10^{-9} $& $7.91347\times10^{-9} $\\
$0.2 $&$2.67755\times10^{-8} $ &$5.67893\times10^{-9} $& $1.51087\times10^{-7} $\\
$0.4 $&$ 5.48083\times10^{-7}$ &$3.22788\times10^{-9} $& $4.65999\times10^{-7} $\\
$0.6 $&$3.94648\times10^{-6} $ &$1.26519\times10^{-8} $& $5.64987\times10^{-7 }$\\
$0.8 $&$0.0000194632$   &$4.30961\times10^{-8} $& $3.83177\times10^{-7 }$\\
$1.0 $&$0.0000764243 $  &$8.41884\times10^{-8} $& $3.73303\times10^{-7 }$\\
\hline
 \end{tabular}
 }
\end{table}

\begin{table}
\caption{ The residual errors $R_1, R_2, R_3$ for different values
of $c$ and $N=15$. }\label{t2}
 \centering
\scalebox{0.9}{
\begin{tabular}{|c|c|c|c|}
  \hline
$c$ & $R_1(\mathcal{T}_{15}, \mathcal{I}_{15}, \mathcal{V}_{15}; \hbar = -1)$  &  $R_2(\mathcal{T}_{15}, \mathcal{I}_{15}, \mathcal{V}_{15}; \hbar = -0.8)$   &  $R_3(\mathcal{T}_{15}, \mathcal{I}_{15}, \mathcal{V}_{15}; \hbar = -0.8)$   \\
    \hline
$0.0$&$ 7.15256\times10^{-7} $ &$1.00187\times10^{-12} $& $3.31113\times10^{-12}$\\
$0.2 $&$4.76837\times10^{-7} $ &$1.28466\times10^{-11}$& $2.76486\times10^{-10}$\\
$0.4 $&$9.53674\times10^{-7}$ &$1.24601\times10^{-10}$& $3.15595\times10^{-9}$\\
$0.6 $&$1.43051\times10^{-6}$ &$1.49157\times10^{-10}$& $1.14451\times10^{-8}$\\
$0.8 $&$3.33786\times10^{-6}$   &$2.91038\times10^{-10}$& $1.74332\times10^{-8}$\\
$1.0 $&$1.90735\times10^{-6}$  &$1.52795\times10^{-9}$& $8.58563\times10^{-9}$\\
\hline
 \end{tabular}
 }
\end{table}

\begin{landscape}

\begin{table}
\caption{ The numerical results of residual error $R_1$ for
different values of $\hbar$ and $N=5, 10, 15$.}\label{t3}
 \centering
\scalebox{0.8}{
\begin{tabular}{|c|c|c|c|c|c|c|c|}
  \hline
$N$&$c$ & $\hbar = -1.2$ & $\hbar = -1.1$  &  $\hbar = -1$   &  $\hbar = -0.9$ & $\hbar = -0.8$  & $\hbar = -0.7$ \\
\hline
   &$0.0$&$2.3936$&$0.2473$&$4.54747\times10^{-13}$&$0.1573$&$0.953603$&$2.31402$\\
   &$0.2$&$2.86231$&$0.393719$&$0.00192233$&$0.193756$&$0.97476$&$2.30218$\\
$5$&$0.4$&$3.3422$&$0.546942$&$0.00769465$&$0.228277$&$0.995169$&$2.29027$\\
   &$0.6$&$3.83332$&$0.706994$&$0.0173277$&$0.260861$&$1.01483$&$2.27832$\\
   &$0.8$&$4.33573$&$0.873909$&$0.0308357$&$0.291503$&$1.03374$&$2.2663$\\
   &$1.0$&$4.84949$&$1.04772$&$0.0482368$&$0.320195$&$1.0519$&$2.25423$\\
\hline
   &$0.0$&$0.00337716$&$0.0000109961$&$0$           &$7.19819\times10^{-6}$&$0.00143155$&$0.0274381$\\
   &$0.2$&$0.00981886$&$0.000116083$&$2.67755\times10^{-8}$&$0.0000356712$&$0.00218667$&$0.0314455$\\
$10$&$0.4$&$0.019624$&$0.00036842$&$5.48083\times10^{-7}$  &$0.0000815957$&$0.00301206$&$0.0354763$\\
   &$0.6$&$0.0335876$&$0.000853229$&$3.94648\times10^{-6}$ &$0.000146411$&$0.00390505$&$0.0395269$\\
   &$0.8$&$0.052855$&$0.00172094$&$0.0000194632$    &$0.000231558$&$0.00486306$&$0.0435937$\\
   &$1.0$&$0.0792877$&$0.00327132$&$0.0000764243$   &$0.000339417$&$0.00588353$&$0.0476735$\\
 \hline
   &$0.0$&$0.0000190735$&$5.72205\times10^{-6}$&$7.15256\times10^{-7}$&$0$&$1.08033\times10^{-6}$&$0.000158743$\\
   &$0.2$&$0.0000305176$&$8.58307\times10^{-6}$&$4.76837\times10^{-7}$&$5.96046\times10^{-8}$&$3.23355\times10^{-6}$&$0.000240277$\\
$15$&$0.4$&$0.0000839233$&$7.62939\times10^{-6}$&$9.53674\times10^{-7}$&$0$&$6.4075\times10^{-6}$&$0.000333367$\\
   &$0.6$&$0.000236511$&$7.62939\times10^{-6}$&$1.43051\times10^{-6}$&$8.34465\times10^{-7}$&$0.0000109375$&$0.000438217$\\
   &$0.8$&$0.00088501$&$0.000038147$&$3.33786\times10^{-6}$&$0$&$0.0000168085$&$0.000554932$\\
   &$1.0$&$0.00294495$&$0.000110626$&$1.90735\times10^{-6}$&$2.38419\times10^{-7}$&$0.0000242442$&$0.000683663$\\
 \hline
 \end{tabular}
 }
\end{table}
\end{landscape}

\begin{landscape}
\begin{table}
\caption{ The numerical results of residual error $R_2$ for
different values of $\hbar$ and $N=5, 10, 15$.}\label{t4}
 \centering
\scalebox{0.8}{
\begin{tabular}{|c|c|c|c|c|c|c|c|c|c|c|}
  \hline
$N$&$c$ & $\hbar = -1$ & $\hbar = -0.9$  &  $\hbar = -0.8$   &  $\hbar = -0.7$ & $\hbar = -0.6$  & $\hbar = -0.5$ & $\hbar = -0.4$ & $\hbar = -0.3$ & $\hbar = -0.2$ \\
\hline
   &$0.0$&$8.67362\times10^{-19}$&$1.53\times10^{-7}$   &$8.96002\times10^{-7}$  &$2.07912\times10^{-6}$  &$3.0741\times10^{-6}$  &$3.14453\times10^{-6}$&$1.84893\times10^{-6}$&$4.6405\times10^{-7}$&$1.94852\times10^{-6}$\\
   &$0.2$&$2.36183\times10^{-6}$&$3.59603\times10^{-7}$ &$1.83824\times10^{-8}$  &$7.03787\times10^{-7}$  &$1.91511\times10^{-6}$ &$2.88405\times10^{-6}$&$2.59665\times10^{-6}$&$6.24361\times10^{-7}$&$1.72845\times10^{-6}$\\
$5$&$0.4$&$0.0000158382$       &$3.24029\times10^{-6}$  &$1.66805\times10^{-7}$  &$2.7399\times10^{-7}$   &$9.90816\times10^{-7}$ &$2.20749\times10^{-6}$&$2.75449\times10^{-6}$&$1.41204\times10^{-6}$&$1.3795\times10^{-6}$\\
   &$0.6$&$0.0000526939$       &$0.000013953$           &$1.50589\times10^{-6}$  &$5.86228\times10^{-7}$  &$6.33103\times10^{-7}$ &$1.45376\times10^{-6}$&$2.49182\times10^{-6}$&$1.9137\times10^{-6}$&$9.44909\times10^{-7}$\\
   &$0.8$&$0.000128933$        &$0.0000400706$          &$6.95057\times10^{-6}$  &$9.10652\times10^{-7}$  &$9.55113\times10^{-7}$ &$8.90192\times10^{-7}$&$1.96673\times10^{-6}$&$2.14988\times10^{-6}$&$4.6329\times10^{-7}$\\
   &$1.0$&$0.000264526$        &$0.000091402$           &$0.0000206229$          &$3.85551\times10^{-8}$  &$1.83977\times10^{-6}$ &$7.08762\times10^{-7}$&$1.32488\times10^{-6}$&$2.14656\times10^{-6}$&$3.12971\times10^{-8}$\\
\hline
   &$0.0$&$8.88178\times10^{-16}$&$7.11076\times10^{-12}$&$1.39264\times10^{-9}$&$2.61784\times10^{-8}$&$1.8088\times10^{-7}$&$6.83613\times10^{-7}$&$1.61316\times10^{-6}$&$2.26423\times10^{-6}$&$9.01673\times10^{-7}$\\
   &$0.2$&$1.16239\times10^{-9}$&$2.9645\times10^{-10}$&$5.67893\times10^{-9}$&$3.01391\times10^{-8}$&$3.37891\times10^{-8}$&$1.58873\times10^{-7}$&$8.46232\times10^{-7}$&$1.87794\times10^{-6}$&$1.5073\times10^{-6}$\\
$10$&$0.4$&$9.90321\times10^{-8}$&$3.27009\times10^{-9}$&$3.22788\times10^{-9}$&$6.49721\times10^{-8}$&$1.43943\times10^{-7}$&$7.98264\times10^{-8}$&$3.8955\times10^{-7}$&$1.43543\times10^{-6}$&$1.77257\times10^{-6}$\\
   &$0.6$&$1.49498\times10^{-6}$&$1.84515\times10^{-8}$&$1.26519\times10^{-8}$&$8.15485\times10^{-8}$&$2.30731\times10^{-7}$&$1.98717\times10^{-7}$&$1.77066\times10^{-7}$&$1.06296\times10^{-6}$&$1.8029\times10^{-6}$\\
   &$0.8$&$0.0000108751$&$2.1865\times10^{-7}$&$4.30961\times10^{-8}$&$8.28309\times10^{-8}$&$3.10485\times10^{-7}$&$2.9468\times10^{-7}$&$1.19973\times10^{-7}$&$8.20621\times10^{-7}$&$1.68352\times10^{-6}$\\
   &$1.0$&$0.0000525467$&$1.77104\times10^{-6}$&$8.41884\times10^{-8}$&$7.48912\times10^{-8}$&$3.8044\times10^{-7}$&$4.06603\times10^{-7}$&$1.34283\times10^{-7}$&$7.22977\times10^{-7}$&$1.48205\times10^{-6}$\\
\hline
   &$0.0$&$1.59162\times10^{-12}$&$1.13687\times10^{-13}$&$1.00187\times10^{-12}$&$1.54\times10^{-10}$&$4.59024\times10^{-9}$&$5.49317\times10^{-8}$&$3.44807\times10^{-7}$&$1.22125\times10^{-6}$&$2.00388\times10^{-6}$\\
   &$0.2$&$1.45519\times10^{-11}$&$3.63798\times10^{-12}$&$1.28466\times10^{-11}$&$4.77598\times10^{-10}$&$6.29957\times10^{-9}$&$1.96057\times10^{-8}$&$5.14657\times10^{-8}$&$6.0141\times10^{-7}$&$1.74667\times10^{-6}$\\
$15$&$0.4$&$8.73115\times10^{-11}$&$1.45519\times10^{-11}$&$1.24601\times10^{-10}$&$4.50314\times10^{-10}$&$1.02114\times10^{-8}$&$5.85214\times10^{-8}$&$8.89341\times10^{-8}$&$2.45201\times10^{-7}$&$1.41322\times10^{-6}$\\
   &$0.6$&$5.41331\times10^{-9}$&$4.36557\times10^{-11}$&$1.49157\times10^{-10}$&$3.06818\times10^{-9}$&$6.67617\times10^{-9}$&$8.18413\times10^{-8}$&$1.69766\times10^{-7}$&$6.71273\times10^{-8}$&$1.10776\times10^{-6}$\\
   &$0.8$&$1.16648\times10^{-7}$&$1.16415\times10^{-10}$&$2.91038\times10^{-10}$&$6.66932\times10^{-9}$&$3.37855\times10^{-9}$&$9.46372\times10^{-8}$&$2.34552\times10^{-7}$&$9.33337\times10^{-9}$&$8.82618\times10^{-7}$\\
   &$1.0$&$1.27265\times10^{-6}$&$1.39698\times10^{-9}$&$1.52795\times10^{-9}$&$1.04319\times10^{-8}$&$1.89696\times10^{-8}$&$9.87109\times10^{-8}$&$2.98313\times10^{-7}$&$4.19607\times10^{-8}$&$7.55504\times10^{-7}$\\
 \hline
 \end{tabular}
 }
\end{table}
\end{landscape}

\begin{landscape}
\begin{table}
\caption{ The numerical results of residual error $R_3$ for
different values of $\hbar$ and $N=5, 10, 15$.}\label{t5}
 \centering
\scalebox{0.7}{
\begin{tabular}{|c|c|c|c|c|c|c|c|c|c|c|c|}
  \hline
$N$&$c$ & $\hbar = -1$ & $\hbar = -0.9$  &  $\hbar = -0.8$   &  $\hbar = -0.7$ & $\hbar = -0.6$  & $\hbar = -0.5$ & $\hbar = -0.4$ & $\hbar = -0.3$ & $\hbar = -0.2$ & $\hbar = -0.1$\\
\hline
   &$0.0$&$8.67362\times10^{-18}$&$8.724\times10^{-7}$&$0.0000135168$&$0.0000597131$&$0.000156055$&$0.000296227$&$0.000433237$&$0.000461549$&$0.000198946$&$0.000632101$\\
   &$0.2$&$0.000129472$&$3.88679\times10^{-6}$&$0.0000232311$&$0.0000367519$&$0.0000380235$&$0.0000855291$&$0.000223225$&$0.000400997$&$0.000394736$&$0.000274581$\\
$5$&$0.4$&$0.00135862$&$0.000177259$&$7.51081\times10^{-6}$&$0.0000564347$&$0.0000381277$&$0.0000198935$&$0.0000296976$&$0.000260365$&$0.000477599$&$0.0000232641$\\
   &$0.6$&$0.00583374$&$0.00124183$&$0.0000647749$&$0.0000352512$&$0.0000876229$&$0.0000240495$&$0.000107956$&$0.0000835355$&$0.000471531$&$0.000266178$\\
   &$0.8$&$0.017109$&$0.0046054$&$0.000574488$&$0.0000322082$&$0.000113154$&$0.0000442726$&$0.00016974$&$0.0000946753$&$0.000398158$&$0.000458718$\\
   &$1.0$&$0.0404312$&$0.0125307$&$0.00229039$&$0.0000254373$&$0.0000588742$&$0.000140393$&$0.000152141$&$0.000247825$&$0.000276823$&$0.00060526$\\
\hline
   &$0.0$&$8.88178\times10^{-16}$&$1.28284\times10^{-11}$&$7.91347\times10^{-9}$&$2.85089\times10^{-7}$&$3.30616\times10^{-6}$&$0.0000202734$&$0.0000802982$&$0.000220035$&$0.000385465$&$0.0000947121$\\
   &$0.2$&$4.92029\times10^{-8}$&$4.72683\times10^{-9}$&$1.51087\times10^{-7}$&$2.10356\times10^{-8}$&$3.15451\times10^{-6}$&$0.000011795$&$0.0000284061$&$0.0000862038$&$0.000263189$&$0.000290276$\\
$10$&$0.4$&$5.84654\times10^{-6}$&$6.25106\times10^{-8}$&$4.65999\times10^{-7}$&$1.54108\times10^{-6}$&$4.21028\times10^{-7}$&$0.0000109938$&$0.0000191687$&$0.0000153365$&$0.000131322$&$0.000390522$\\
   &$0.6$&$0.000109276$&$1.85419\times10^{-7}$&$5.64987\times10^{-7}$&$3.14461\times10^{-6}$&$2.74231\times10^{-6}$&$8.67918\times10^{-6}$&$0.0000254225$&$6.65097\times10^{-6}$&$0.0000152451$&$0.000417307$\\
   &$0.8$&$0.00093273$&$0.0000128647$&$3.83177\times10^{-7}$&$4.83092\times10^{-6}$&$5.14217\times10^{-6}$&$4.8722\times10^{-6}$&$0.0000315151$&$2.60466\times10^{-6}$&$0.0000712265$&$0.000389391$\\
   &$1.0$&$0.00512464$&$0.000139456$&$3.73303\times10^{-7}$&$6.37062\times10^{-6}$&$7.49196\times10^{-6}$&$2.3893\times10^{-6}$&$0.0000314601$&$0.000026181$&$0.00012284$&$0.000322735$\\
\hline
   &$0.0$&$2.27374\times10^{-13}$&$2.27374\times10^{-13}$&$3.31113\times10^{-12}$&$9.86057\times10^{-10}$&$4.98968\times10^{-8}$&$9.59473\times10^{-7}$&$9.74223\times10^{-6}
$&$0.0000603884$&$0.000229812$&$0.000330239$\\
   &$0.2$&$0$             &$2.72848\times10^{-12}$&$2.76486\times10^{-10}$&$1.5668\times10^{-8}
$&$4.30215\times10^{-8}$&$7.57963\times10^{-7}$&$5.88135\times10^{-6}$&$0.0000229083$&$0.000103157$&$0.000342435$\\
$15$&$0.4$&$3.8126\times10^{-9}$&$1.09139\times10^{-10}$&$3.15595\times10^{-9}$&$3.00904\times10^{-8}$&$3.66815\times10^{-7}$&$5.20925\times10^{-7}$&$4.20601\times10^{-6}$&$0.0000143819
$&$0.0000268466$&$0.000289918$\\
   &$0.6$&$3.14903\times10^{-7}$&$1.01863\times10^{-9}$&$1.14451\times10^{-8}$&$1.30044\times10^{-8}$&$5.98062\times10^{-7}$&$1.85369\times10^{-6}$&$2.01119\times10^{-6}$&$0.0000163409$&$7.71528\times10^{-6}$&$0.000201946$\\
   &$0.8$&$7.62218\times10^{-6}$&$2.2701\times10^{-9}$&$1.74332\times10^{-8}$&$1.26489\times10^{-7}$&$6.84324\times10^{-7}$&$2.97462\times10^{-6}$&$4.24188\times10^{-7}$&$0.0000192142$&$0.0000125839$&$0.000100126$\\
   &$1.0$&$0.0000949907$&$5.54137\times10^{-8}$&$8.58563\times10^{-9}$&$3.07671\times10^{-7}$&$6.09615\times10^{-7}$&$3.99947\times10^{-6}$&$2.38387\times10^{-6}$&$0.000019579$&$5.39521\times10^{-9}$&$2.04689\times10^{-7}$\\
 \hline
 \end{tabular}
 }
\end{table}
\end{landscape}

\section{Conclusion}
The mathematical model of HIV infection for CD4$^+$T cells is an
applicable and robust model to analyze, track and control the
infection. In this study, the HATM was introduced by combining the
Laplace transformations and the HAM to approximate the presented
model. The obtained solution of HATM depends on auxiliary parameters
and functions specially the convergence control parameter $\hbar$.
It is useful tool to identify and control the region of convergence.
Existence of auxiliary functions and parameters are the advantages
of the HATM which transformed the presented method to the flexible
and applicable scheme than the other semi-analytical methods.
Furthermore, the convergence theorem was proved to support the HATM
analytically. The rate of convergence depends on parameter $\hbar$.
In order to show the regions of convergence some $\hbar$-curves were
plotted for $N=5, 10, 15$. The efficiency and accuracy of method
were shown by applying the errors of residual functions. The
numerical errors based on the residual functions were obtained for
different values of convergence control parameter $\hbar$ and number
of iteration $N$.

\end{document}